\documentstyle[12pt]{article}
\begin{document}
\begin{titlepage} 

\title{The energy-momentum of plane-fronted gravitational waves
in the teleparallel equivalent of GR}

\author{J. W. Maluf$\,^{*}$ and S. C. Ulhoa\\
Instituto de F\'{\i}sica, \\
Universidade de Bras\'{\i}lia\\
C. P. 04385 \\
70.919-970 Bras\'{\i}lia DF, Brazil\\}
\date{}
\maketitle

\begin{abstract}
We show that in the framework of the teleparallel equivalent of
general relativity the gravitational energy-momentum of plane-fronted
gravitational waves contained in an arbitrary three-dimensional
volume $V$ may be easily obtained and is nonpositive in 
the frame of static observers.
\end{abstract}
\thispagestyle{empty}
\vfill
\par
\bigskip
\noindent (*) e-mail: wadih@unb.br\par
\end{titlepage}
\newpage

\noindent

\section{Introduction}
The observation of gravitational waves is presently one of the main 
challanges of gravitational physics. Significant efforts are being 
undertaken to construct the necessary apparatus for the detection
of the waves \cite{AB,EF}. On the theoretical side there is no general
agreement regarding the description of the energy-momentum carried by
gravitational waves. The reason is that there is no generally 
accepted definition for the gravitational energy-momentum. The use of
noncovariant energy-momentum pseudotensors totally obscures the 
analysis of the issue. In the framework of pseudotensors one 
concludes that the energy carried by a gravitational wave, 
considered as a gravitational fluctuation of the spacetime, is not 
gauge invariant, i.e., is coordinate dependent \cite{PM}. 

The energy-momentum of gravitational waves has been investigated 
with the help of the Bel-Robinson tensor \cite{BR}. The idea is to
consider a gravitoelectromagnetic stress-energy tensor
\cite{Mashhoon1}, whose properties are similar to
the Maxwell stress-energy tensor in electrodynamics. Covariant
approaches to the energy-momentum of gravitational waves have
been investigated \cite{Mashhoon2,DT}. However, it must be
noted that the Bel-Robinson tensor requires an additional
multiplicative factor with physical dimensions to relate it to an 
acceptable energy-momentum tensor. 

In this article we will show that in the spacetime of plane-fronted
gravitational waves the energy-momentum definition established in the 
teleparallel equivalent of general relativity (TEGR) allows obtaining
the gravitational energy-momentum enclosed in an arbitrary volume $V$ 
of the three-dimensional space in an easy way. Moreover it is 
straightforward to verify that the resulting expression is 
nonpositive for any $V$. The resulting expression for the energy is 
naturally invariant under coordinate transformations of the 
three-dimensional space, and under time reparametrizations. In this 
sense, it is gauge invariant.

Notation: space-time indices $\mu, \nu, ...$ and SO(3,1)
indices $a, b, ...$ run from 0 to 3. Time and space indices are
indicated according to
$\mu=0,i,\;\;a=(0),(i)$. The tetrad field is denoted $e^a\,_\mu$,
and the torsion tensor reads
$T_{a\mu\nu}=\partial_\mu e_{a\nu}-\partial_\nu e_{a\mu}$.
The flat, Minkowski space-time metric tensor raises and lowers
tetrad indices and is fixed by
$\eta_{ab}=e_{a\mu} e_{b\nu}g^{\mu\nu}= (-+++)$. The determinant of 
the tetrad field is represented by $e=\det(e^a\,_\mu)$.\par

\section{The energy-momentum definition in the Lagrangian framework}

We assume that the spacetime geometry is defined by the
tetrad field $e^a\,_\mu$ only. In this case we note that the only
possible nontrivial definition for the torsion tensor is given by
$T_{a\mu\nu}=\partial_\mu e_{a\nu}-\partial_\nu e_{a\mu}$.
This torsion tensor is related to the antisymmetric part of Cartan's
connection 
$\Gamma^\lambda_{\mu\nu}=e^{a\lambda}\partial_\mu e_{a\nu}$, which is
the connection of the Weitzenb\"ock spacetime. However, the tetrad
field also yields the metric tensor, which establishes the
Riemannian geometry. Therefore in the framework of 
a geometrical theory based only on 
the tetrad field one may use the concepts of both Riemannian and
Weitzenb\"ock geometries. $T_{a\mu\nu}$ is not covariant under local 
Lorentz transformations. The tetrad fields are interpreted as 
reference frames adapted to preferred fields of observers in
spacetime. This interpretation is possible by identifying the 
$e_{(0)}\,^\mu$ components of the frame with the four-velocities
$u^\mu$ of the observers, $e_{(0)}\,^\mu=u^\mu$ \cite{Maluf4}.
Therefore two different sets tetrad fields that yield the same metric 
tensor, and which are related by a local Lorentz transformation,
represent different frames in the same spacetime.

The equivalence of the TEGR with Einstein's general relativity may be
understood by means of an identity between the scalar curvature 
$R(e)$ constructed out of the tetrad field and a combination of
quadratic terms of the torsion tensor, 

\begin{equation}
eR(e)\equiv -e({1\over 4}T^{abc}T_{abc}+{1\over 2}T^{abc}
T_{bac}-T^aT_a)
+2\partial_\mu(eT^\mu)\,.
\label{1}
\end{equation}
The formulation of Einstein's general relativity in the context of the
teleparallel geometry is presented in review articles 
\cite{Hehl,Hehl1,FG} and in recent books \cite{MB,TO}.

The Lagrangian density of the TEGR is given by 
the combination of the quadratic terms on the right hand side of 
eq. (1),

\begin{eqnarray}
L&=& -k e({1\over 4}T^{abc}T_{abc}+{1\over 2}T^{abc}T_{bac}-
T^aT_a) -L_M\nonumber \\
&\equiv& -ke\Sigma^{abc}T_{abc}-L_M\,, 
\label{2}
\end{eqnarray}
where $k=c^3/16\pi G$, $T_a=T^b\,_{ba}$, 
$T_{abc}=e_b\,^\mu e_c\,^\nu T_{a\mu\nu}$ and

\begin{equation}
\Sigma^{abc}={1\over 4} (T^{abc}+T^{bac}-T^{cab})
+{1\over 2}( \eta^{ac}T^b-\eta^{ab}T^c)\;.
\label{3}
\end{equation}
$L_M$ stands for the Lagrangian density for the matter fields.
The field equations derived from (2) are equivalent to 
Einstein's equations. They read

\begin{equation}
e_{a\lambda}e_{b\mu}\partial_\nu (e\Sigma^{b\lambda \nu} )-
e (\Sigma^{b\nu}\,_aT_{b\nu\mu}-
{1\over 4}e_{a\mu}T_{bcd}\Sigma^{bcd} )={1\over {4k}}eT_{a\mu}\,,
\label{4}
\end{equation}
where
$\delta L_M / \delta e^{a\mu}=eT_{a\mu}$. 
It is possible to show that that the left hand side of the equation
above may be rewritten as
${1\over 2}e\left[ R_{a\mu}(e)-{1\over 2}e_{a\mu}R(e)\right]$.

Equation (4) may be simplified as 

\begin{equation}
\partial_\nu(e\Sigma^{a\lambda\nu})={1\over {4k}}
e\, e^a\,_\mu( t^{\lambda \mu} + T^{\lambda \mu})\;,
\label{5}
\end{equation}
where $T^{\lambda\mu}=e_a\,^{\lambda}T^{a\mu}$ and
$t^{\lambda\mu}$ is defined by

\begin{equation}
t^{\lambda \mu}=k(4\Sigma^{bc\lambda}T_{bc}\,^\mu-
g^{\lambda \mu}\Sigma^{bcd}T_{bcd})\,.
\label{6}
\end{equation}
In view of the antisymmetry property 
$\Sigma^{a\mu\nu}=-\Sigma^{a\nu\mu}$ it follows that

\begin{equation}
\partial_\lambda
\left[e\, e^a\,_\mu( t^{\lambda \mu} + T^{\lambda \mu})\right]=0\,.
\label{7}
\end{equation}
The equation above yields the continuity (or balance) equation,

\begin{equation}
{d\over {dt}} \int_V d^3x\,e\,e^a\,_\mu (t^{0\mu} +T^{0\mu})
=-\oint_S dS_j\,
\left[e\,e^a\,_\mu (t^{j\mu} +T^{j\mu})\right]\,.
\label{8}
\end{equation}
We identify
$t^{\lambda\mu}$ as the gravitational energy-momentum tensor
\cite{Maluf1}, and

\begin{equation}
P^a=\int_V d^3x\,e\,e^a\,_\mu (t^{0\mu} 
+T^{0\mu})\,,
\label{9}
\end{equation}
as the total energy-momentum contained within a volume $V$ of the
three-dimensional space.
In view of (5), eq. (9) may be written as 

\begin{equation}
P^a=-\int_V d^3x \partial_j \Pi^{aj}\,,
\label{10}
\end{equation}
where $\Pi^{aj}=-4ke\,\Sigma^{a0j}$. $\Pi^{aj}$ is the momentum
canonically conjugated to $e_{aj}$ \cite{Maluf5}.
The expression above is the 
definition for the gravitational energy-momentum presented in ref.
\cite{Maluf2}, obtained in the framework of the Hamiltonian vacuum 
field equations. We note that (7) is a true energy-momentum 
conservation equation.

By transforming the SO(3,1) index $a$ in eq. (4) into a spacetime
index $\lambda$, say, we find that the left hand side of (4) becomes
the symmetric tensor density ${1\over 2}eR_{\lambda\mu}$. For 
spin $0$ and spin $1$ fields the energy-momentum tensor for the 
matter fields $T_{\lambda\mu}$ is symmetric. For Dirac spinor fields
$T_{\lambda\mu}$ is also symmetric provided the coupling of the spinor
field with the gravitational field is constructed out of the 
Levi-Civita connection

$$^o\omega_{\mu ab}=-{1\over 2}e^c\,_\mu(
\Omega_{abc}-\Omega_{bac}-\Omega_{cab})\;,$$

$$\Omega_{abc}=e_{a\nu}(e_b\,^\mu\partial_\mu e_c\,^\nu-
e_c\,^\mu\partial_\mu e_b\,^\nu)\;.$$
In the evaluation of $T_{\lambda\mu}$ one first obtains a tensor that
is not symmetric. However, it has been shown that anti-symmetric part
of $T_{\lambda\mu}$ vanishes in view of the Dirac equation. This 
issue has been discussed in detail in ref. \cite{Maluf6} and
references therein. As for the tensor $t^{\lambda\mu}$, in general it
is not symmetric. So far it is not clear the meaning of the 
anti-symmetric part of $t^{\lambda\mu}$.

The emergence of a nontrivial total divergence is a feature of 
theories with torsion. The 
integration of this total divergence yields a surface integral. If we 
consider the $a=(0)$ component of eq. (10) and adopt asymptotic 
boundary conditions for the tetrad field we find \cite{Maluf2} that 
the resulting 
expression is precisely the surface integral at infinity that
defines the ADM energy \cite{ADM}. This fact is a strong indication 
(but no proof) that eq. (10) does indeed represent the gravitational
energy-momentum.

The evaluation of definition (10) is carried out in the
configuration space.
The definition is invariant under (i) general coordinate
transformations of the three-dimensional space, (ii) time 
reparametrizations, and (iii) global SO(3,1) transformations. The
non-invariance of eq. (10) under the local SO(3,1) group
reflects the frame dependence of the definition. We have argued
\cite{Maluf4} that this dependence is a natural feature 
of $P^a$, since in the TEGR each set of tetrad fields is interpreted
as a reference frame in spacetime. 

It is worthwhile to recall a simple physical situation in which the
frame dependence of the gravitational energy-momentum takes place.
For this purpose we consider a black hole of mass $m$ and an
observer that is very distant from the black hole. The black hole
will appear to this observer as a particle of mass $m$, with energy
$P^{(0)}=mc^2$ ($m$ is the 
rest mass of the black hole, i.e., the mass of the black hole
in the frame where the black hole is at rest).
If, however, the black hole is
moving at velocity $v$ with respect to the observer, then its total
gravitational energy will be $P^{(0)}=\gamma m c^2$, where 
$\gamma=(1-v^2/c^2)^{-1/2}$. This example is a consequence of
the special theory of relativity, and demonstrates the 
frame dependence of the gravitational energy-momentum. We note that
the frame dependence is not restricted to observers at spacelike 
infinity. It holds for observers located 
everywhere in space.

\section{The energy-momentum of plane-fronted gravitatonal waves}

Definition (10) for the gravitational energy-momentum may be easily
applied to plane-fronted gravitational waves. We can show that
these waves carry negative gravitational energy.

A plane-fronted gravitational wave is an exact solution of 
Einstein's equations. A wave that travels along the $z$ direction 
may be described by the line element \cite{HS}

\begin{equation}
ds^2=dx^2+dy^2+2du\,dv+H(x,y,u)du^2\,,
\label{11}
\end{equation}
where the function $H(x,y,u)$ satisfies

\begin{equation}
\biggl( {\partial^2 \over {\partial x^2}}+
{\partial^2 \over {\partial y^2}}\biggr)H(x,y,u)=0\,.
\label{12}
\end{equation}
Transforming the $(u,v)$ to $(t,z)$ coordinates, where

$$u={1\over \sqrt{2}}(z-t)\,,$$

$$v={1\over \sqrt{2}}(z+t)\,,$$
we find

\begin{equation}
ds^2=\biggl({H\over 2} -1\biggr)dt^2+dx^2+dy^2+
\biggl({H\over 2}+1\biggr) dz^2-H\,dt dz\,.
\label{13}
\end{equation}
The function $H$ is only required to satisfy (12). However,
it would be interesting to specify $H$ such that it describes 
a wave-packet  \cite{HS}. The inverse metric tensor reads

\begin{equation}
g^{\mu\nu}=\pmatrix{
-{1\over 2}H-1&0&0&-{1\over 2}H \cr
0&1&0&0\cr
0&0&1&0\cr
-{1\over 2}H&0&0&-{1\over 2}H+1}\,.
\label{14}
\end{equation}
 
We will choose the tetrad field that is adapted to static
observers in spacetime. Therefore the tetrad field must satisfy
$e_{(0)}\,^i=0$. One suitable construction, adapted to the symmetry 
of the gravitational field, is

\begin{equation}
e_{a\mu}=\pmatrix{
-A&0&0&-B\cr
0&1&0&0\cr
0&0&1&0\cr
0&0&0&C}\,,
\label{15}
\end{equation}
where

\begin{eqnarray}
A&=&\biggl(-{H\over 2}+1\biggr)^{1/2}\,, \nonumber \\
AB&=&{H\over 2}\,, \nonumber \\
AC&=&1 \,.
\label{16}
\end{eqnarray}
In (15) $a$ and $\mu$ label rows and columns, respectively. The
geometrical interpretation of (15) is best understood if we consider
the inverse components $e_a\,^\mu$. We verify that

\begin{equation}
e_{(0)}\,^\mu=(1/A,0,0,0)\,,
\label{17}
\end{equation}
and

\begin{eqnarray}
e_{(1)}\,^\mu&=& (0,1,0,0)\,,  \nonumber \\
e_{(2)}\,^\mu&=& (0,0,1,0)\,,  \nonumber \\
e_{(3)}\,^\mu&=& (\,-H/(2A)\,,0,0,A). 
\label{18}
\end{eqnarray}
Note that if $H<<1$ we have $A\cong 1-H/4$ and therefore
$e_{(3)}\,^i=(0,0,A)\cong (0,0,1-H/4)$.

The frame is determined by fixing six conditions on $e_{a\mu}$. 
Equation (17) fixes the kinematic state of the frame, since
the three velocity conditions $e_{(0)}\,^i=0$ ensure that the frame 
is static. Three other conditions fix the spatial sector of the
frame. According to (18), $e_{(1)}\,^\mu$, $e_{(2)}\,^\mu$ and
$e_{(3)}\,^\mu$ are unit vectors along the $x$, $y$ and $z$ axis,
respectively. Therefore (14) fixes the orientation of the frame
in spacetime: the $e_{(3)}\,^\mu$ component is oriented along the
direction of propagation of the wave.
Alternatively, the frame may be determined by fixing
the six components of the acceleration tensor $\phi_{ab}=-\phi_{ba}$
\cite{Maluf4}. These are the inertial accelerations (translational 
and rotational) that are necessary to maintain the frame in
a given inertial state in spacetime. However in the present case the
characterization of the frame by means of (17) and (18) seems to be
more appropriate. Finally it must be noted that by requiring $H=0$ we
obtain $e_a\,^\mu=\delta_a^\mu$, and consequently $T_{a\mu\nu}=0$.

The nonvanishing components of $T_{\mu\nu\lambda}$ are

\begin{eqnarray}
T_{001}&=& {1\over 2} \partial_1 A^2 \nonumber \\
T_{002}&=& {1\over 2} \partial_2 A^2 \nonumber \\
T_{003}&=& {1\over 2} \partial_3 A^2 - A\partial_0 B \nonumber \\
T_{013}&=& -A\partial_1 B \nonumber \\
T_{023}&=& -A\partial_2 B \nonumber \\
T_{301}&=& B\partial_1 A \nonumber \\
T_{302}&=& B\partial_2 A \nonumber \\
T_{303}&=& B\partial_3 A +{1\over 2}
\partial_0(-B^2+C^2) \nonumber \\
T_{313}&=& {1\over 2} \partial_1(-B^2+C^2) \nonumber \\
T_{323}&=& {1\over 2} \partial_2(-B^2+C^2)\,.
\label{19}
\end{eqnarray}
In the expressions above we have $(-B^2+C^2)=g_{33}$. 

In order to calculate the gravitational energy-momentum
we find it more convenient to transform expression (10) 
into a surface integral,

\begin{equation}
P^a=4k\,\oint_S dS_i (e\Sigma^{a0i})\,.
\label{20}
\end{equation}
The determinant $e$ is simply $e=AC=1$.
After long but straightforward calculations we obtain

\begin{eqnarray}
dS_i(e\Sigma^{(0)0i})&=&-{1\over {8(-g_{00})^{1/2}}}
\biggl[dy\,dz\,\partial_1 H+dz\,dx\,\partial_2 H\biggr]
\,, \nonumber \\
dS_i(e\Sigma^{(1)0i})&=&{1\over {8(-g_{00})}}
\biggl[dy\,dz\,\partial_0 H+dx\,dy\,\partial_1 H\biggr]
\,, \nonumber \\
dS_i(e\Sigma^{(2)0i})&=&{1\over {8(-g_{00})}}
\biggl[dz\,dx\,\partial_0 H+dx\,dy\,\partial_2 H\biggr]
\,, \nonumber \\
dS_i(e\Sigma^{(3)0i})&=&-{1\over {8(-g_{00})^{1/2}}}
\biggl[dy\,dz\,\partial_1 H+dz\,dx\,\partial_2 H\biggr]
\,.
\label{21}
\end{eqnarray}
In the evaluation of $\Sigma^{(1)0i}$ and $\Sigma^{(2)0i}$ we have
used the relation $\partial_3 g_{00}=-\partial_0 g_{00}$, since
the metric quantities in (11) are functions of $u$. 

We may now return to volume integrals. We have

\begin{equation}
\oint_S dS_i (e\Sigma^{(0)0i})=
\int_V d^3x \partial_i(e\Sigma^{(0)0i})=-{1\over 8}
\int_V d^3x\,\partial_i\biggl[ {1\over {(-g_{00})^{1/2}}}
\partial_i H \biggr]\,.
\label{22}
\end{equation}
Taking into account that 

$$\partial_i(\partial_i H)=
\biggl( {\partial^2 \over {\partial x^2}}+
{\partial^2 \over {\partial y^2}}\biggr)H=0\,,$$
we find

$$\oint_S dS_i (e\Sigma^{(0)0i})=-{1\over {32}}\int_V
{{(\partial_i H)^2}\over {(-g_{00})^{3/2}}}\,.$$
Therefore we obtain

\begin{equation}
P^{(0)}=P^{(3)}=-{k\over 8}\int_V d^3x
{{(\partial_i H)^2}\over {(-g_{00})^{3/2}}} \leq 0\,,
\label{23}
\end{equation}
assuming that $(-g_{00})>0$.
Next we calculate $P^{(1)}$. In view of (21) we have 

\begin{eqnarray}
\oint_S dS_i(e\Sigma^{(1)0i})&=&
\int_V d^3x\,\partial_i (e\Sigma^{(1)0i})\nonumber \\
{}&=& \int_V d^3x \biggl[
\partial_1\biggl(  {1\over {8(-g_{00})}}\partial_0 H \biggr)+
\partial_3 \biggl( {1\over {8(-g_{00})}}\partial_1 H \biggr)
\biggr]\nonumber \\
{}&=&{1\over 8}\int_V \biggl[
{1\over {(-g_{00})}}\partial_1 \partial_0 H -
{1\over {(-g_{00})^2}}\partial_0 H\partial_1 \biggl(-{H\over 2}
\biggr) \nonumber \\
{}&{}&+
{1\over {(-g_{00})}}\partial_3 \partial_1 H -
{1\over {(-g_{00})^2}}\partial_1 H\partial_3 \biggl(-{H\over 2}
\biggr) \biggr]\,.
\label{24}
\end{eqnarray}
Considering that $\partial_0 H=-\partial_3 H$ we see that the
expression above vanishes. The same result holds in the calculation
of $P^{(2)}$. Therefore we conclude that

\begin{equation}
P^{(1)}=P^{(2)}=0\,.
\label{25}
\end{equation}
Equations (25) and (23) imply 

\begin{equation}
P^aP^b\eta_{ab}=0\,.
\label{26}
\end{equation}

\section{Final remarks}

Equation (26) is a feature of a plane electromagnetic wave. 
If the similarity between plane gravitational and electromagnetic
waves holds, we may assert that the plane-fronted gravitational
wave transports the excitations of a massless field. Moreover in view
of eq. (23) we see that the energy enclosed by an arbitrary volume 
$V$ is nonpositive. Therefore the energy carried by a plane-fronted
gravitational wave is negatively defined and gauge invariant, i.e.,
coordinate independent (in addition we note that eq. (26) is 
invariant under global SO(3,1) transformations of the frame). 
Therefore we conclude that the noncovariant character of the energy
carried by gravitational waves \cite{PM} is restricted to the 
investigation in the context of pseudotensors.

The analysis above shows that in the framework of the TEGR definition
(10) allows a satisfactory treatment of the energy-momentum of 
plane-fronted waves. In fact the application of eq. (10) to any
configuration of gravitational field is conceptually simple. One has
to specify the frame adapted to a field of observers in 
spacetime endowed with velocities $u^\mu=e_{(0)}\,^\mu$. The 
orientation of the frame in the three-dimensional space may be 
fixed according to the symmetry of the gravitational field. This 
choice of the tetrad field is not arbitrary. It is determined by
the choice of the reference frame adapted to a field of observers
in spacetime, exactly like in the special theory of relativity or
in the Newtonian limit of general relativity.

\bigskip
\noindent {\bf Acknowledgement}\par
\noindent This work was supported in part by CNPQ.

\end{document}